\documentclass[fleqn,12pt,a4paper]{article}
\usepackage{amsmath}
\usepackage{amsfonts}
\usepackage{fancyheadings}
\newtheorem{theorem}{Theorem}[section]
\numberwithin{equation}{section} % uses amsmath or older amstex package
\newcommand{\R}{\mathbb{R}} % uses amsfonts or amstex package
\newcommand{\N}{\mathbb{N}} % uses amsfonts or amstex package
\newcommand{\1}{\mathbf{1}}
\newcommand{\A}{\mathbf{A}}
\newcommand{\bx}{\mathbf{x}}
\newcommand{\bp}{\mathbf{p}}
\newcommand{\bv}{\mathbf{v}}
\newcommand{\bn}{\mathbf{0}}
\newcommand{\om}{\boldsymbol{\omega}}
\newcommand{\pb}{\mathbf{\bar{p}}}
\newcommand{\balpha}{\boldsymbol{\alpha}}
\DeclareMathOperator{\Div}{div} % uses amsmath package
\DeclareMathOperator{\const}{const} % uses amsmath package
\DeclareMathOperator*{\supp}{supp} % uses amsmath package
\DeclareMathOperator*{\slim}{s-lim} % uses amsmath package
\newcommand{\spmlim}{\slim_{\substack{t_{_{+}}\to\infty\\t_{_{-}}\to-\infty}}}
\newcommand{\pblim}{\xrightarrow[|\pb | \to \infty ]{}} % uses amsmath package
\newcommand{\beql}[1]{\begin{equation} \label{#1}}
\begin{document}
\pagestyle{fancy}
\headrulewidth=0pt
\lhead{V.~Enss}
\chead{\thepage}
\rhead{Inverse Scattering}
\lfoot{}
\cfoot{}
\rfoot{}
\thispagestyle{empty}
\begin{center}
{\Large A New Look at the Multidimensional\\ Inverse Scattering
Problem\footnote{To appear in: \textit{Understanding Physics},
A.K.~Richter ed., Copernicus Gesellschaft, Katlenburg-Lindau, 1998, pp. 31--;
ISBN 3-9804862-2-2 (Proceedings Bonn 1996).}\\[3ex]
Volker Enss}\\[3ex]
Institut f\"ur Reine und Angewandte Mathematik, RWTH
Aachen\\ 
D-52056 Aachen, Germany\\[2ex]
email: enss@rwth-aachen.de,\quad
http://www.iram.rwth-aachen.de/$\sim$enss/
\end{center}
\vspace{2ex}

\begin{abstract}
%
%    Text of Abstract Abstract Abstract Abstract 00000000000000000
%
As a prototype of an evolution equation we consider the Schr\"odinger
equation
$$
i (d/dt) \Psi (t) = H \: \Psi (t), \quad  H = H_0 + V(\bx)
$$
for the Hilbert space valued function
$\Psi(\cdot ): \; \R \to {\cal H} =  L^2 (\R^\nu )\;$
which describes the state of the system at time $t \,$ in space dimension
$\nu \geq 2$. The kinetic energy operator
$\, H_0 \,$ may be $\, H_0 = - (1/2m)\, \Delta \,$\vspace{0.3ex}
(nonrelativistic quantum mechanics for a particle of mass $m$),
$\, H_0 = \sqrt{-\Delta + m^2} \,$ (relativistic kinematics, Klein-Gordon
equation), the Dirac operator, or ..., while
the potential $\, V(\bx) \to 0 \,$ suitably as $\,|\bx| \to \infty$.

We present a \textit{geometrical} approach to the inverse scattering
problem. For given scattering operator $\,S\,$ we show uniqueness of
the potential, we give explicit limits of the high-energy behavior of
the scattering operator, and we give reconstruction formulas for the
potential.

Our mathematical proofs closely follow physical intuition. A key
observation is that at high energies translation of wave packets
dominates over spreading during the interaction time.
Extensions of the method cover e.g.\ Schr\"odinger
operators with magnetic fields, multiparticle systems, and wave equations.
\end{abstract}

% section 1111111111111111111

\section{Introduction, the Schr\"odinger Equation}
The {\bfseries Schr\"odinger equation} is a linear evolution equation for a 
function of time $\,t \in \R\,$ with values in a state space (phase space)
$\,{\cal H}\,$ which is a Hilbert space:
\begin{displaymath}
\Psi(\cdot) : \R \to {\cal H}.
\end{displaymath}
The initial value problem reads
\beql{1.1}
i\frac{d}{dt}\, \Psi (t) = H\,\Psi (t),\quad \Psi (0) = \Psi,
\end{equation}
with a linear operator $\,H\,$ acting on $\,{\cal H}$. (We use units where
Planck's constant $\hbar =1$.) This type of equation 
includes as special cases nonrelativistic and relativistic quantum mechanics,
the Dirac equation, the linear wave equation (with the usual method to
transform a second order equation into a first order system), and other
evolution equations.  Splitting off the factor $\,i\,$ is just for
convenience because in important applications $\,H\,$ is symmetric.
Since the operator $\,H\,$ is typically unbounded, care is needed to ensure
solvability of the equation. In the models mentioned above the operator
$\,H\,$ is self-adjoint on a suitably chosen domain $\,{\cal D}(H)$.
Then Stone's theorem (or the spectral theorem and functional calculus,
see e.g.\ \cite{RS2}\,) ensure that the exponential $\,\exp\{-itH\}\,$
is a well defined unitary operator for all $\,t\in\R\:$ and that 
\beql{1.2}
\Psi(t) = e^{-itH} \Psi
\end{equation}
is the unique global solution of the initial value problem (\ref{1.1}).

In the following we will describe our geometrical approach to the inverse
problem for the Schr\"odinger equation as an equation which describes
the motion of particles according to the laws of quantum mechanics.
We will exploit physical intuition to help us solve the mathematical
problems. The methods and results carry over to other evolution equations
as well.

We will discuss the differences of nonrelativistic and relativistic
kinematics as far as they are relevant here. The time scales for
interaction and for spreading of wave functions differ at high
energies. This implies the simplicity of the leading behavior of the
scattering operator because only the translational part of the time
evolution matters as long as the interaction is strong. We obtain
explicit formulas for the high energy scattering operator which can
be used to reconstruct the potential uniquely. In the present paper
we want to explain \textit{why} the statements are true and how
physical intuition and mathematical proofs are closely analogous.
While we give all major steps of the proofs for two typical examples
we refer to the papers for some more technical estimates and further
examples.

\textit{Acknowledgement.\;} This paper is dedicated to Wolfgang Kundt
who has strongly influenced my view of science.
Part of the material was prepared during my stay at the Institute for
Advanced Study, Pinceton, NJ, USA. I gratefully acknowledge the hospitality
and support.

% section 222222222222

\section{Particles in Quantum Mechanics}
We describe the state of a quantum mechanical particle in $\,\nu$-dimensional
space by a normalized vector $\,\Psi \in {\cal H}$.
(Due to the superposition principle we may restrict ourselves to pure states
and choose a normalized vector instead of the equivalence class of vectors
differing by a global phase factor.)
The vector can conveniently be represented in
various ways (similar to basis changes in linear algebra), e.g.\ by a
square integrable function $\,\psi (\cdot ) \in L^2 (\R^\nu ,\, dx)\,$ with
volume measure $\,dx$. Instead of a function depending on the
\textit{position} or \textit{configuration space variable}
$\,\bx \in \R^\nu \,$ one can  use its Fourier transform
\beql{2.1}
\hat{\psi}(\cdot) \in L^2 (\R^\nu ,\, dp), \quad
\hat{\psi}(\bp) := (2\pi)^{-\nu/2} \int dx \; e^{-i\bp \bx}\; \psi(\bx)
\end{equation}
depending on the \textit{momentum variable} $\,\bp \in \R^\nu \,$
with normalization
\begin{displaymath}
\| \Psi\|^2 = \int dx\; |\psi (\bx)|^2 = \int dp\; |\hat{\psi}(\bp)|^2 = 1.
\end{displaymath}
We use for the abstract state vector a capital letter $\,\Psi$,  for
its representation as a function of position $\,\psi(\bx)$, or its
momentum space wave function
$\,\hat{\psi}(\bp)$,
respectively, and write
\begin{alignat}{5}
&{\cal H} &&\;\longleftrightarrow\; &\; L^2 ( & \R^\nu ,\, dx)\; &
&\;\longleftrightarrow\; 
& \; L^2 ( & \R^\nu ,\, dp) \notag \\
&\Psi &&\;\longleftrightarrow\; && \psi(\bx) &
&\;\longleftrightarrow\; 
& & \hat{\psi}(\bp)  \label{2.2}
\end{alignat}
to indicate the switching between representations. 

For a given state $\,\Psi\,$ the probability measures $\,\mu_\bx\,$ on
configuration space and $\,\mu_\bp\,$ on momentum space, respectively,
\beql{2.3}
\mu_\bx (A) = \int_A dx\; |\psi (\bx)|^2 \quad \text{ and } \quad
\mu_\bp (B) = \int_B dp\; |\hat{\psi}(\bp)|^2 
\end{equation}
describe the probabilities to find the particle in the 
(Lebesgue measurable) subset $\,A \subset \R^\nu\,$ of configuration space 
or $\,B \subset \R^\nu\,$ of momentum space.
One may visualize such a state as a cloud of very many particles where
$\,\mu_\bx (A)\,$ describes the fraction of them which have their position
in $\,A\,$ and, similarly, $\,\mu_\bp (B)\,$ is the fraction with momentum
in $\,B$. Such a state is also called a \textit{wave packet}.
Classical point particles with a $\,\delta$-like
distribution in position and/or momentum space are impossible in quantum
mechanics. The state space does not contain such idealized objects,
in agreement with observations.

We extend the triple of representations of state vectors to the linear
operators acting on them.
\begin{center}
\parbox{6.9em}{Abstract operator acting on ${\cal H}$}
$\;\longleftrightarrow\;$
\parbox{10.2em}{Action on configuration space wave functions} 
$\;\longleftrightarrow\;$
\parbox{9.4em}{Action on momentum space wave functions}
\vspace{1ex}
\end{center}

The Fourier transformation (\ref{2.1}) interchanges
differentiation and multiplication of a function with its argument.
Thus we obtain for the position and momentum operators, respectively,
\begin{alignat}{5}
&\bx &&\;\;\longleftrightarrow\;\; &&\bx &&\;\;\longleftrightarrow\;\; 
& \;\;i &\nabla_\bp\, ,
\label{2.4}\\[1ex]
&\bp &&\;\;\longleftrightarrow\;\; &\;\; - i &\nabla_\bx\; 
&&\;\;\longleftrightarrow\;\; &&\bp.
\label{2.5}
\end{alignat}
In our notation sometimes we do not distinguish between the abstract
operator on $\,{\cal H}\,$ and its representaion as a multiplication
operator on the corresponding space on which it is ``diagonal''.

If the forces acting on the particle are described as the negative gradient 
of a potential function $\,V(\bx)\,$ (conservative mechanical system) then the
generator $\,H\,$ of the time evolution, the \textit{Hamiltonian} or
\textit{Schr\"odinger operator}, is the energy operator
\beql{2.6}
H = H_0 + V(\bx)
\end{equation}
which is a sum of the kinetic energy operator $\,H_0$ -- responsible for the 
kinematics -- and the real valued potential energy which determines 
the dynamics. The
potential should decrease suitably as $\,|\bx| \to \infty$, see Section
\ref{SecDynamics} for precise conditions on $\,V$.

% section 333333333333333

\section{Kinematics} \label{SecKinematics}
The kinetic energy operator or \textit{free Hamiltonian} $\,H_0\,$
usually is a function $\,H_0(\bp)\,$ of the momentum of the particle.
We will study two typical cases, nonrelativistic (NR) and relativistic (Rel)
kinematics.
In the first case
\beql{3.1}
\text{NR:} \quad H_0(\bp) = \frac{1}{2m} \bp^2.
\end{equation}
It acts as a multiplication operator on $\,\hat{\phi}\,$ and as a differential
operator on $\,\phi\,$:
$$
H_0\,\Phi \;\;\longleftrightarrow\;\; (H_0\,\phi)(\bx) = 
- \frac{1}{2m} (\Delta \phi)(\bx)
\;\;\longleftrightarrow\;\; H_0(\bp) \, \hat{\phi}(\bp)= 
\frac{1}{2m} \bp^2\,\hat{\phi}(\bp).
$$
Generally, the \textit{velocity operator} is the change of position in time:
\beql{3.2}
\bv(\bp) = \frac{d}{dt}\, e^{itH_0} \,\bx\,
e^{-itH_0} \Bigr|_{t=0}
= i \, [H_0,\, \bx] = \nabla_\bp\, H_0(\bp)\, , 
\end{equation}
a function of the momentum operator.
In the nonrelativistic case it is
\beql{3.3}
\text{NR:} \quad \bv(\bp) = \frac{\bp}{m}\, .
\end{equation}
Note that the velocity is unbounded, the speed tends to
infinity if the kinetic energy or the momentum does so.

Let us now turn to the relativistic case corresponding e.g.\ to the 
Klein-Gordon equation. 
\beql{3.4}
\text{Rel:} \quad H_0(\bp) = \sqrt{\bp^2 c^2 + m^2 c^4} = \sqrt{\bp^2 + m^2}
\end{equation}
if we use units of measurement
such that the speed of light $\,c=1$. Here the
velocity operator is
\beql{3.5}
\text{Rel:} \quad \bv(\bp) =\nabla_\bp\, H_0(\bp) = 
c \,\frac{\bp c}{\sqrt{\bp^2 c^2 + m^2 c^4}} =
\frac{\bp}{\sqrt{\bp^2 + m^2}}\, .
\end{equation}
In this case the speed is bounded by $\,c=1\,$ (the speed of light).

The {\bfseries free time evolution} operator is a simple multiplication
operator in momentum space [\,and a complicated oscillating convolution
operator in configuration space\,]
\beql{3.6}
e^{-i t H_0} \Phi \;\;\longleftrightarrow\;\; (e^{-i t H_0} \phi)(\bx) 
\;\;\longleftrightarrow\;\; e^{- i t H_0(\bp)} \hat{\phi}(\bp).
\end{equation}
While for short times the free classical and quantum time evolutions differ 
considerably they behave similarly for large times. Asymptotically, the 
distribution in configuration space of a quantum state is in good
approximation the same as that of the corresponding cloud of free classical
particles, of the ``classical wave packet''.  For later applications
we study a particular family of states $\,\Phi_\pb\,$ with compact
momentum support around a very large average momentum
$\,\pb\in \R^{\nu}\,$. The operator $\,\exp(i\pb \bx)$, a unitary function
of the position operator $\,\bx$, shifts a state in momentum space by $\,\pb$:
\begin{alignat}{5}
&\Phi_\bn & &\longleftrightarrow & &\phi_\bn(\cdot) &
&\longleftrightarrow & \hat{\phi}_\bn(\cdot) &\in C_0^{\infty}(\R^\nu)
\label{3.7}\\[1.5ex]
\Phi_\pb = &e^{i \pb \bx} \Phi_\bn & &\longleftrightarrow &
\;\phi_\pb(\bx) &= e^{i \pb \bx} \phi_\bn(\bx) &
&\longleftrightarrow & \;\hat{\phi}_\pb(\bp) &=\hat{\phi}_\bn(\bp-\pb).
\label{3.8}
\end{alignat}
Since $\phi_\bn(\cdot) \in {\cal S}(\R^\nu)$, the Schwartz space of
rapidly decreasing functions, these states are well localized in 
configuration space, too, uniformly in $\pb$. They have average velocities
around $\,\bv (\pb)\in\R^{\nu}$, where
\beql{3.9}
\bv (\pb) = \nabla H_0(\pb) =: v(\pb) \; \om =
\begin{cases}
\pb/m & \text{NR}\, ,\\[1ex]
\pb/\sqrt{\pb^2 + m^2} & \text{Rel}\, .
\end{cases}
\end{equation}
In the given examples the unit direction vector $\,\om = \bv(\pb)/|\bv (\pb)| 
= \pb / |\pb|\,$ of the velocity $\,\bv (\pb)\,$ has the same direction as
$\,\pb\,$ and the speed $\,v(\pb) = |\bv (\pb)| \,$ depends on
$\,|\pb|\,$ only.

Consider in the following only ``large'' $\,\pb\,$ such that  the minimal 
velocity in the support of $\,\hat{\phi}_\pb\,$ is at least
$\,2/3^{\text{rd}}\,$ of the average velocity:
\beql{3.10}
\inf\left\{ \frac{v(\bp)}{v(\pb)}\biggm| \bp \in 
\supp \hat{\phi}_\pb\right\} \geq \frac{2}{3}\, .
\end{equation}
In our examples e.g.\ $\,|\pb| \geq 3 \sup\{\, |\bp|\,\mid \,\bp \in \supp 
\hat{\phi}_\bn\}\,$ will do. As we are mainly interested in the
high-energy behavior $\,|\pb|\to\infty\,$ this restriction (\ref{3.10})
is harmless, it excludes components in the state with particles of
low or zero velocities which would require special treatment below.

In our context we have to control the localization in configuration
space of freely evolving wave packets. This depends mainly on the
support of the state in velocity (momentum) space. Therefore, we have
chosen compactly supported momentum space wave functions.
Then in configuration space the states cannot have compact support as well 
but rapid falloff is sufficient there.
A special case of such {\bfseries propagation properties} (or
non--propagation) of quantum wave packets for long times
\cite{Enss:Prop}, \cite{RS3}, is
\beql{3.11}
\int\limits_{|\bx|< t\, v(\pb)/2} dx\;
|(e^{-i t H_0} \phi_\pb)(\bx)|^2
< \frac{\const(\Phi_\bn,\,n)}{(1+|t\, v(\pb)|)^n}
\end{equation}
for any $\,n \in \N\,$ \textit{uniformly} for large $\,\pb$.
A classical free particle which starts at time $\,0\,$ from the origin
and has momentum $\,\bp\in \supp \hat{\phi}_\pb\,$ will be localized
at time $\,t\,$ in the region
\beql{3.12}
\bx(t)\in\left\{\bx= t\,\bv(\bp)\Bigm| \bp\in \supp \hat{\phi}_\pb \right\}
\subset \left\{\,\bx\, \mid \,|\bx - t\, \bv(\pb)|<v(\pb)/3 \right\}.
\end{equation}
The ``classically forbidden'' region $\,|\bx|< t\, v(\pb)/2\,$ is
separated from the ``allowed region'' by at least $\, t\, v(\pb)/6$.
The state mainly propagates within the classically allowed region which
moves away from the origin with a positive minimal speed.
The ``quantum tails'' of the wave packet in the classically forbidden region
do not vanish, nevertheless, they decay very fast in time, both in the future
and past. This is physically and mathematically in close analogy to rays
versus waves in optics. While the shadow behind an obstacle is not totally
black due to diffraction it is, nevertheless, quite dark away from the region
which can be reached by straight rays (the role of the increasing
separation $\, t\, v(\pb)/6$\,).  We will need below only the estimate of
non--propagation (\ref{3.11}). It is proved with a stationary phase 
estimate of an integral with rapidly oscillating integrand.

Kinematics has a strong influence on the high-energy behavior of the
scattering strength. Wave packets will be influenced by the potential
only as long as they are mainly localized in the region where the potential
is strong. Due to the propagation property
this is essentially a finite time interval, the interaction time, 
proportional to $\,v(\pb)^{-1}$. Thus the strength of the 
interaction (which is proportional to the duration of the interaction)
will vanish as $\,|\pb| \to \infty\,$ in the nonrelativistic case
and it will have a finite limit in the relativistic case. This is a crucial
difference of the two cases.

% section 444444444444444444444

\section{Dynamics} \label{SecDynamics}
Let us now turn to the interacting (perturbed) time evolution
\beql{4.1}
e^{-itH} \Psi, \quad H = H_0 + V(\bx).
\end{equation}
The free Hamiltonian $\,H_0\,$ is self-adjoint on its domain
\begin{displaymath}
{\cal D}(H_0) = \{\Psi \in {\cal H}\mid \hat{\phi}(\pb) \text{ and }
H_0(\bp)\,\hat{\phi}(\pb) \in L^2(\R^{\nu},\,dp)\}.
\end{displaymath}
The configuration
space wave functions lie in the Sobolev spaces $\,W^{2,2}(\R^{\nu})\,$
in the nonrelativistic case and in $\,W^{1,2}(\R^{\nu})\,$ in the relativistic
case, respectively.
First we will consider \textit{short-range} potentials $\,V(\bx) =
V^s(\bx)\,$ which are roughly those which decrease at least like
$\,|\bx|^{-(1+\varepsilon)}$, $\,\varepsilon > 0$,
as $\,|\bx| \to \infty$. More precisely, the set of short-range potentials is
\beql{4.2}
\mathcal{V}^s = \biggl\{V^s \; \biggm|  \int_0 ^{\infty} 
\sup_{ |\bx| \geq R}|V^s (\bx)| \;\; dR  < \infty \biggr\}.
\end{equation}
For simplicity of presentation we will restrict ourselves in this paper
to bounded potentials. The sum is self-adjoint on the domain $\,{\cal D}(H)
={\cal D}(H_0)$. Singular potentials can easily be included using standard
techniques \cite{RS2}. This covers the physically relevant local
singularities like $\, 1/|\bx|\,$
for Coulomb or Yukawa potentials. One simply has to regularize the
singularities by a free resolvent or by a function $\,f(\bp)\,$ of the 
momentum operator which has compact support and is the identity on
$\,\Phi_\bn\,$. In addition, with suitable adjustments we will treat
long-range potentials $\,V^\ell \,$ which decrease towards infinity
e.g.\ as slowly as the Coulomb potential, see Section
\ref{SecHighEnergy2}. Then
\beql{4.3}
V(\bx) = V^s(\bx) + V^{\ell} (\bx), \quad V^s \in \mathcal{V}^s.
\end{equation}

In the present context a short-range potential behaves similarly to a 
compactly supported one. Depending on the required accuracy it is essentially
concentrated in a ball of some radius $\,R\,$ around the origin. 

The influence on the particle by the force $\,-\nabla V(\bx)\,$
is relevant only as long as the particle is essentially localized in
the interaction region, i.e.\ where
the potential is strong. Correspondingly, the state space can be
split into two orthogonal components. The bound states remain forever under 
the influence of the potential, they constitute the \textit{pure point}
spectral subspace which is spanned by the eigenvectors of $\,H$. These
states remain localized uniformly in time. Orthogonal to these are the
scattering states which form the \textit{continuous} spectral subspace
$\, \mathcal{H}^{\text{cont}} (H) = \{\text{eigenvectors of } H\}^\perp$, 
they leave the interaction region for
large times (see e.g.\ Theorem XI.115 in \cite{RS3}\,).
The latter are studied in scattering theory.

% section 5555555555555555555555

\section{Scattering} \label{SecScattering}
As a general reference for mathematical scattering theory see e.g.\
\cite{RS3} or \cite{Thaller} for the Dirac equation. It is well known that
for short-range potentials ($H=H_0 + V^s$) the asymptotic motion of
scattering states is an essentially free motion: For any
scattering state $\,\Psi \in \mathcal{H}^{\text{cont}} (H)\,$ there exist
free asymptotic configurations $\, \Phi^\pm \in \mathcal{H}\,$ such that
\beql{5.1}
\left\| e^{-it\,[H_0 + V^s]}\, \Psi -  
e^{-itH_0}\, \Phi^\pm \right\| \to 0 \quad \text{as }\;
t \to \pm \infty.
\end{equation}
This is usually called asymptotic completeness of the wave operators.
Similarly, for any \textit{incoming} configuration $\,\Phi^-\,$ or 
\textit{outgoing} $\,\Phi^+\,$ there is a corresponding state
$\,\Psi \in \mathcal{H}^{\text{cont}} (H)\,$ such that (\ref{5.1}) holds
(existence of wave operators). These statements can be proved for
any dimension $\,\nu\,$ of configuration space using 
propagation estimates like (\ref{3.11}). The proof of (\ref{7.2})
below is similar.

A convenient tool to describe scattering is the scattering operator
$\,S\,$ which maps an incoming configuration $\,\Phi^-\,$ to the corresponding  
outgoing configuration $\,\Phi^+\,$ of the same state $\,\Psi$. For given
$\,\Phi^-\,$ let
\begin{align*}
&\Psi = \lim_{t_{_-} \to -\infty}
e^{i\, t_{_-} [H_0 + V^s]} \; e^{-i\, t_{_-}\, H_0}\;\Phi^-\quad
\text{ and}
\\
&\Phi^+= \lim_{t_{_+} \to \infty}e^{i\,t_{_+} H_0} \; e^{-i\,
t_{_+} [H_0 + V^s]} \; \Psi.
\end{align*}
Then 
\begin{align} \label{5.2}
&S(t_{_+},t_{_-}) := 
e^{i\,t_{_+} H_0} \; e^{-i\, t_{_+} [H_0 + V^s]} \;
e^{i\, t_{_-} [H_0 + V^s]} \; e^{-i\, t_{_-}\, H_0}\; ,
\\[2ex]
&S := \spmlim 
\; S(t_{_+},t_{_-}), \quad\text{satisfies}\quad S \, \Phi^- = \Phi^+.
\label{5.3} \end{align}
For microscopic particles for which quantum mechanics is an adequate
description one cannot really observe more details of the scattering
process than those encoded in the scattering operator. We denote the
mapping
\beql{5.4}
\mathcal{V}^s \to L(\mathcal{H}), \quad V^s \mapsto S = S(V^s)
\end{equation}
as the \textit{scattering map} from short-range potentials to bounded 
(unitary) scattering operators on the Hilbert space of asymptotic 
configurations.

The {\bfseries direct} problem of scattering theory is to determine for a
given potential $\,V\,$ the scattering operator while the {\bfseries inverse}
problem is to determine the potential(s) if the scattering operator or
part of it is known.

% section 66666666666666666666

\section{Uniqueness of the Potential}
We denote by $\,F(H_0 \geq E)\,$ the multiplication operator in
momentum space with the characteristic function of the set
$\,\{\bp\in\R^{\nu} \mid H_0(\bp)\geq E\}$, i.e.\ the spectral
projection of the kinetic energy operator to energies above $\,E$.
The main results about uniqueness are of the following form.
They are a corollary of the asymptotic behavior of the scattering operator
shown below.
\begin{theorem}
The scattering map $\,S\colon \mathcal{V}^s \to L(\mathcal{H})\,$ is
injective. Actually, the high-energy part of the scattering operator
alone: $\:S\; F(H_0 \geq E), \; E\,$ arbitrarily large, determines the 
short-range potential uniquely.
\end{theorem}

In the nonrelativistic case similar results 
go back to Faddeev (1956) and Berezanski (1958), the
strongest results were by Saito (1984), all using time-independent
methods. Our geometrical time-dependent proof covers a wider class of
potentials and, more importantly, it is simpler. It has been extended 
to show analogous results for long-range potentials, magnetic fields, 
multiparticle systems, for the Dirac equation, and other systems, see Sections
\ref{SecHighEnergy1}, \ref{SecHighEnergy2}, and \ref{SecReconstruction}.

% section 777777777777777777777

\section{Time Scales for Interaction and Spreading}  \label{SecTimeScales}
For high energy states as constructed in (\ref{3.8}) scattering theory becomes
simple because two time scales, a short interaction time $\,T_I(\pb)\,$
and a longer kinematical time of spreading $\,T_{Sp}(\pb)\,$ 
satisfy $\,T_I(\pb) / T_{Sp}(\pb) \to 0\,$ as $\,|\pb| \to \infty$.
For a potential which is essentially supported in a ball of radius
$\,R\,$ the \textbf{interaction time} is of the order
$\,T_I(\pb) = R/ v(\pb)$. Due to propagation like (\ref{3.11}) we have
\beql{7.1}
S(t_{_+},t_{_-})\; \Phi_\pb^- \approx S \; \Phi_\pb^- \quad
\text{if both} \quad \pm t_\pm \gg T_I(\pb).
\end{equation}
More precisely, for $\,\Phi_\pb ,\;\Phi'_\pb\,$ as in (\ref{3.8}) and 
any $\,\varepsilon > 0\,$ there is a radius $\,\rho(\varepsilon)\,$
such that uniformly for large $|\pb|$ (which satisfy (\ref{3.10})\,)
\beql{7.2}
| ( \Phi'_\pb,\;[\,S\; - S(t_{_+},t_{_-})\,] \; \Phi_\pb )| 
< \frac{\varepsilon}{v(\pb)}\quad 
\text{ if } \quad \pm t_\pm > \rho(\varepsilon) / v(\pb).
\end{equation}
Intuitively, $\,\rho(\varepsilon)\,$ measures the radius of the interaction 
region and the extension in configuration space of the states up to effects
of size $\,\varepsilon$. The bound (\ref{7.2}) is physically
intuitive and it is a crucial estimate which will be used in Section
\ref{SecHighScat} to justify the interchanging of limits.

To prove (\ref{7.2}) one has to bound a term
\begin{align}
&\left\| \lim_{t\to\infty} e^{i\, t [H_0 + V^s]} \; e^{-i\,
t\, H_0}\;\Phi'_\pb -
e^{i\, t_{_+} [H_0 + V^s]} \; e^{-i\, t_{_+}\, H_0}\;\Phi'_\pb
\right\| \notag\\[2ex]
&= \left\| \int_{t_{_+}}^\infty dt\;\frac{d}{dt}\;
e^{i\, t [H_0 + V^s]} \; e^{-i\,t\, H_0}\;\Phi'_\pb \right\| 
\leq \int_{t_{_+}}^\infty dt\left\| V^s\;e^{-i\,t\, H_0}\;\Phi'_\pb
\right\| \label{7.3}
\end{align}
and a similar term with $\,t_{_-}\,$ and $\,\Phi_\pb\,$. The
integrand in(\ref{7.3}) can be split into two terms
\begin{align*}
&\left\| V^s\;e^{-i\,t\, H_0}\;\Phi'_\pb \right\|\\[2ex]
&\leq \| V^s\; F(|\bx|> v(\pb)\,t/2) \| + \| V^s\|\:\left\|F(|\bx|\geq
v(\pb)\,t/2)\;e^{-i\,t\, H_0}\;\Phi'_\pb \right\| \\[2ex]
&=: h_1(v(\pb)\,t) + h_2(v(\pb)\,t)
\end{align*}
where $\,F(\cdot)\,$ here denotes the multiplication operator with
the characteristic function of the indicated region in configuration
space. The functions $\,h_1\,$ and $\,h_2\,$ are integrable
due to (\ref{4.2}) and (\ref{3.11}), respectively. With the new
variable $\,r:= v(\pb)\,t\,$ the integral (\ref{7.3}) is bounded by
\begin{displaymath}
\frac{1}{v(\pb)}\;\int_{r_{_+}}^\infty dr\;[\,h_1(r) + h_2(r)\,]
\leq \frac{\varepsilon}{v(\pb)}
\end{displaymath}
for $\,r_{_+} \geq \rho(\varepsilon)\,$ large enough. This proves
(\ref{7.2}).\vspace{2ex}

The kinematical {\bfseries time scale of spreading} $\,T_{Sp}(\pb)\,$
denotes the time after which spreading of wave packets becomes relevant
in the time evolution. As
\begin{displaymath}
H_0(\bp) \, \Psi_\pb = H_0(\bp) \, e^{i\pb\bx} \, \Psi_\bn 
= e^{i\pb\bx} \, H_0(\pb + \bp) \, \Psi_\bn
\end{displaymath}
we will expand the kinetic energy function around $\,\pb\,$
\beql{7.4}
H_0(\pb + \bp) =: H_0(\pb) + \nabla H_0(\pb) \cdot \bp + H_2(\pb,\: \bp).
\end{equation}
The first summand is a number giving an irrelevant phase, the second equals
$\,\bv(\pb)\cdot \bp\,$ by (\ref{3.9}). It is the
dominant term which -- as a multiple of the momentum operator -- generates 
a translation of the wave packet without changing its shape. 
Only the third term $\, H_2\,$ (which is defined by (\ref{7.4})\,) 
is responsible for the spreading of the wave packet. 
In our examples of ``power like'' Hamiltonians this part of the 
Hamiltonian is weak compared to the translational 
component: On a compact subset of momentum space like 
$\,\bp \in \supp \hat{\phi}_\bn$
\beql{7.5}
\frac{T_I(\pb)}{T_{Sp}(\pb)}\sim\frac{| H_2(\pb,\: \bp)|}{v(\pb)}
\leq \frac{\const}{|\pb|} \pblim 0.
\end{equation}
In the nonrelativistic case we have $\,H_2(\pb,\: \bp) = \bp^2/2m\,$
which is independent of $\,\pb\,$ and bounded on $\,\supp
\hat{\phi}_\bn\,$.
\begin{displaymath}
\frac{| H_2(\pb,\: \bp)|}{v(\pb)} = \frac{\bp^2/2m}{|\pb|/m}
= \frac{\bp^2/2}{|\pb|}\,.
\end{displaymath}
In the relativistic case the denominator is bounded but the numerator
decreases. With the shorthand $\,a:=\sqrt{\pb^2 + m^2}\,$ we get for
$\,|\pb|\,$ large enough
\begin{displaymath}
\frac{| H_2(\pb,\: \bp)|}{v(\pb)} = 
\frac{|a\{\sqrt{1+2\,(\pb\,\bp/a^2) + (\bp^2/a^2)}-1-\pb\,\bp/a^2\}|}{|\pb|/a}
\leq 2 \frac{\bp^2/2}{|\pb|}\,.
\end{displaymath}
This proves (\ref{7.5}).
Therefore, the time $\,T_{Sp}(\pb)\,$ is by a factor proportional to
$\,|\pb|\,$ longer than $\,T_I(\pb)$. For large $\,|\pb|\,$ we may choose
times when the scattering due to the potential is over but the spreading has
not yet really started. Alternatively, we may use radii for this splitting
like the interaction radius $\,\rho(\varepsilon)\,$ of (\ref{7.2}).

Usually, an interacting time evolution is complicated because the
translation of a wave packet, its spreading, and the influence of
the potential all occur at the same time.
In the high-energy limit it is sufficient for the calculation of the 
scattering operator to treat translation of wave packets rather than their
correct free evolution. Since in this limit spreading occurs only
when the interaction is negligible, i.e.\ when the free and interacting time
evolutions are almost the same, the effect of spreading is cancelled
(becomes invisible) in the scattering operator. Thus, high energy scattering
is simple and it can be inverted simply! The motion during the interaction
is dominated by the translational part
which is common to classical and quantum physics. The typical quantum 
effect of spreading of wave functions which results from the absence
of localized states with sharp momentum is of lower order.

% section 88888888888888888888

\section{High Energy Scattering} \label{SecHighScat}
The crucial uniformity of the estimate (\ref{7.2}) enables us to
interchange the limits
\\
$\pm t_\pm \to \infty\,$ and $\,|\pb| \to \infty$.
This simplifies the remaining discussion very much. Actually, as to be
expected, not the time but the separation from the region of a strong
potential determines the quality of approximation. With correspondingly
chosen variables
\\$r_\pm := t_\pm\, v(\pb)\,$ we have
\begin{align}
&\lim_{|\pb| \to \infty}( \Phi'_\pb,\; S\,\Phi_\pb ) =
\lim_{|\pb| \to \infty} \; \lim_{\pm t_\pm \to \infty}
( \Phi'_\pb,\; S(t_{_+},t_{_-})\,\Phi_\pb ) \notag \\[1ex]
&= \lim_{|\pb| \to \infty} \; \lim_{\pm r_\pm \to \infty} 
\left( \Phi'_\pb,\; S\left(\frac{r_{_+}}{v(\pb)},\,\frac{r_{_-}}{v(\pb)}
\right) \Phi_\pb \right) \notag \\[1ex]
&= \lim_{\pm r_\pm \to \infty} \; \lim_{|\pb| \to \infty}
\left( \Phi'_\pb,\; S\left(\frac{r_{_+}}{v(\pb)},\,\frac{r_{_-}}{v(\pb)}
\right) \Phi_\pb \right). 
\label{8.1}
\end{align}
As seen in (\ref{7.2}) the asymptotic equality (\ref{8.1}) remains true
even after multiplication with $\,v(\pb)\,$ which is a much 
stronger statement in the nonrelativistic case. To determine
\beql{8.2}
( \Phi'_\pb,\; S(t_{_+},t_{_-})\,\Phi_\pb )
= \left( \Phi'_\bn,\;e^{-i\pb\bx}\, S(t_{_+},t_{_-})\, 
e^{i\pb\bx}\,\Phi_\bn \right)
\end{equation}
for large finite times and $\,\pb\,$ consider e.g.\ the second pair of
factors in (\ref{5.2}).
\begin{align}
& e^{-i\pb\bx} \; e^{it_{_-} [H_0 + V^s]}\; e^{-it_{_-} H_0}\; e^{i\pb\bx}
\notag \\[1ex]
&= e^{it_{_-} [H_0(\bp + \pb) + V^s(\bx)]}\; e^{-it_{_-} H_0(\bp + \pb)}\;
\notag \\[1ex]
&=  
e^{it_{_-} [H_0(\pb) + \bv(\pb)\cdot \bp + H_2(\pb, \bp) + V^s(\bx)]}\;
e^{-it_{_-} [H_0(\pb) + \bv(\pb)\cdot \bp + H_2(\pb, \bp)]} \notag 
\\[1ex]
&= e^{it_{_-} [\bv(\pb)\cdot \bp + H_2(\pb, \bp) + V^s(\bx)]}\;
e^{-it_{_-} [\bv(\pb)\cdot \bp + H_2(\pb, \bp)]}  \notag \\[1ex]
&= e^{ir_{_-} [\om\cdot \bp + \{H_2(\pb, \bp)/v(\pb)\} 
+ \{V^s(\bx)/v(\pb)\}]}\;
e^{-ir_{_-} [\om\cdot \bp + \{H_2(\pb, \bp)/v(\pb)\}]} \label{8.3}
\end{align}
using again $\,t_\pm = r_\pm / v(\pb)\,$ and the direction $\,\om =
\bv(\pb)/v(\pb)\,$ as in (\ref{3.9}). Due to (\ref{7.5}) the functions
of the momentum operator
\beql{8.4}
[\,\om\cdot \bp + \{H_2(\pb, \bp)/v(\pb)\}\,] \pblim \om\cdot\bp
\end{equation}
converge in strong resolvent sense and similarly for the other exponent.
Therefore, for fixed $\,r_{_-}\,$ and large $\,|\pb|\,$ the following
approximation is good:
\begin{align}
& e^{ir_{_-} [\om\cdot\bp + \{H_2(\pb, \bp)/v(\pb)\} 
+ \{V^s(\bx)/v(\pb)\}]}\;
e^{-ir_{_-} [\om\cdot \bp + \{H_2(\pb, \bp)/v(\pb)\}]} \notag \\[1ex]
&\approx  e^{ir_{_-} [\om\cdot\bp + \{V^s(\bx)/v(\pb)\}]}\;
e^{-ir_{_-} \om\cdot\bp} \label{8.5} \\[1ex]
&= \exp \left\{ \frac{-i}{v(\pb)} \int_{r_{_-}}^0 dr\: 
V^s(\bx + \om\, r) \right\}.
\label{8.6}
\end{align}
The approximation (\ref{8.5}) is the only approximation we have to make!
If $\{H_2(\pb, \bp)/v(\pb)\}$ would commute with
$\,\{V^s(\bx)/v(\pb)\}\,$ then we would have exact cancellation and
(\ref{8.5}) would be an equality as well.
A careful estimate of the correction terms can be given for all
Hamiltonians considered here. It is uniform in $\,r_{_-}\,$ and
when compared to $\,\{V^s(\bx)/v(\pb)\}\,$ it has additional falloff
like $\, 1/|\pb|\,$ for $\pb \to \infty\,$ due to (\ref{7.5}).
For the proofs we refer to the papers cited in the theorems below. 

Equation (\ref{8.6}) is verified easily because as functions of
$\,r_{_-}\,$ both expressions solve the same initial value problem
\begin{displaymath}
\frac{d}{dr_{_-}} A(r_{_-}) = A(r_{_-})\;\, \frac{-i}{v(\pb)}
\, V^s(\bx + \om\, r_{_-}),\quad A(0) = \1.
\end{displaymath}
The same analysis of the first two factors in the expression
(\ref{5.2}) for the scattering operator yields analogously to
(\ref{8.6}) the factor
\begin{displaymath}
\exp \left\{ \frac{-i}{v(\pb)} \int^{\,r_{_+}}_0 dr\:
V^s(\bx + \om\, r) \right\}.
\end{displaymath}
Combining this with (\ref{8.6}) we obtain for large $\,|\pb|\,$ 
\beql{8.7}
\begin{aligned}
( \Phi'_\pb\, , \; S(t_{_+},\, t_{_-}) \, \Phi_\pb )
&=  \left( \Phi'_\bn,\;e^{-i\pb\bx}\, 
S\left(\frac{r_{_+}}{v(\pb)},\,\frac{r_{_-}}{v(\pb)}\right)\, 
e^{i\pb\bx}\,\Phi_\bn \right)
\\[1ex]
&\approx  \left( \Phi'_\bn \, , \;\exp \left\{ \frac{-i}{v(\pb)} 
\int_{r_{_-}}^{\,r_{_+}} dr\; V^s( \bx + \om r)\right\}\;\Phi_\bn \right).
\end{aligned}
\end{equation}
\beql{8.8}
( \Phi'_\pb\, , \; S \: \Phi_\pb )
\approx  \left( \Phi'_\bn \, , \;\exp \left\{ \frac{-i}{v(\pb)} 
\int_{-\infty}^\infty dr\; V^s( \bx + \om r)   \right\}\;\Phi_\bn \right).
\end{equation}
These expressions confirm our intuitive expectation discussed in 
Section \ref{SecKinematics} 
that the limiting behavior of the scattering operator
depends strongly on the growth or boundedness of $\,v(\pb)\,$ as $\,|\pb|\to
\infty$. We will discuss the two cases separately.

% section 99999999999999999999999

\section{High Energy Limits of the Scattering Operator,\\
the Short-Range Case}
\label{SecHighEnergy1}
Next we give the limiting behavior of the scattering operator in several
typical cases, $\,\Phi_\pb$, $\,\Phi'_\pb$, and  $\,\pb\in\R^\nu\:$ as given 
in (\ref{3.8}), (\ref{3.10}). The integrals extend over the real line.
In the quotations we include similar results obtained by other
methods, sometimes under more restrictive assumptions.
\begin{theorem}
{\bf (scalar relativistic, short-range, \cite{Jung})}\\
For the scalar relativistic Hamiltonian
\begin{displaymath}
H = \sqrt{\bp^2 + m^2} + V^s(\bx)
\end{displaymath}
with $v(\pb) \to 1$ one obtains
\beql{9.1}
(\Phi'_\pb , \; S\, \Phi_\pb) 
\pblim \left(\Phi'_\bn , \;
\exp\left\{-i\:\int dr \; V^s(\bx + \om r)\right\} \, \Phi_\bn\right).
\end{equation}
\end{theorem}
\vspace*{4ex}

\noindent
If, however, $\,v(\pb)\to\infty\,$ we can expand the exponential in
(\ref{8.8})
\beql{9.2}
\begin{split}
\exp & \left\{ - \frac{i}{v(\pb)} \int dr\; V^s(\bx + \om r)\right\}\\[1ex]
& \approx 1 - \frac{i}{v(\pb)} \int dr\; V^s(\bx + \om r) + \cdots\\
\end{split}
\end{equation}
which explains the following nonrelativistic result. The leading
behavior of the scattering operator is the identity operator (no
scattering). The next order correction depends on the potential.

\begin{theorem}
{\bf (nonrelativistic, short-range, \cite{Faddeev}, \cite{Berezanskii},
\cite{Saito}, \cite{EnssWeder:Inverse}, \cite{EnssWeder:Geometrical})}\\
For the Hamiltonian
\begin{displaymath}
H = \frac{1}{2m} \bp^2 + V^s(\bx)
\end{displaymath}
\beql{9.3}
v(\pb) \;(\Phi'_\pb , \; i(S-\1)\: \Phi_\pb) \pblim 
\int dr \; (\Phi'_\bn , \; V^s(\bx + \om r) \: \Phi_\bn).
\end{equation}
\end{theorem}

This result is to be expected from the Born approximation. It holds 
also under the given weaker assumptions on the falloff of the potential
where the validity of the Born approximation is not established.

The estimate (\ref{7.2}) and the remark following (\ref{8.6})
justify that multiplication with $\,v(\pb) \sim |\pb|\,$ is
permitted. The approximation is better than $\, V^s / v(\pb)\,$.
\vspace*{1ex}

\noindent{\bfseries Remark}\\
In all these limits there are {\bfseries error bounds} for large but finite 
$|\pb|$ which are explicit. E.g.\ in equation (\ref{9.3}) (and similarly in
(\ref{10.5}) etc.) we obtain
$$
\left| v(\pb) \;(\Phi'_\pb , \; i(S-\1)\, \Phi_\pb) - 
\int\! dr \; (\Phi'_\bn , \, V^s(\bx + \om r) \, \Phi_\bn)\right|
\leq \frac{\const(\Phi'_\bn ,\,\Phi_\bn,\, V^s )}{|\pb|}\,.
$$

\begin{theorem}
{\bf (Dirac equation, \cite{Ito}, \cite{Jung})}\\
For the Dirac Hamiltonian of a particle in a continuous short-range 
electromagnetic field
\begin{displaymath}
H = \balpha\!\cdot\! \bp + \beta m + V(\bx), \quad V(\bx) = 
(A_0 - \balpha\!\cdot\! \A)(\bx)
\end{displaymath}
the high-energy limit of the scattering operator is
\begin{displaymath}
\lim_{|\pb| \to \infty} (\Phi'_\pb , \, S \:\Phi_\pb) =
\!\left(\Phi'_\bn , \;\exp \left\{ -i\!\int\! dr\;
(A_0 - \balpha \!\cdot\! \om \;\; \om\!\cdot\!\A)(\bx + \om r) 
\right\} \Phi_\bn\right)\!.
\end{displaymath}
In the Fouldy-Wouthuysen representation let $\,S_\pm\,$ 
denote the scattering operator on the positive/negative energy subspaces and 
$\,\bx_{_{NW}}\,$ the Newton-Wigner position operator then
$$
\slim_{|\pb| \to \infty} e^{-i\,\pb\bx_{_{NW}}}S_\pm e^{i\,\pb\bx_{_{NW}}} = 
\exp \left\{ -i\!\int\! dr\; (A_0 \mp \balpha\!\cdot\!\om\;\; \om\!\cdot\!\A)
(\bx_{_{NW}} + \om r) \right\}.
$$
\end{theorem}
Similar results hold for other matrix valued potentials $V$.

% section 10 10 10 

\section{High Energy Limits of the Scattering Operator,\\ 
Long-Range Potentials, Magnetic Fields, etc.}
\label{SecHighEnergy2}
The only long-range potential in physics is the Coulomb potential
between electrically charged particles. The mathematical treatment
includes more general potentials which may decrease slower and
need not be centrally symmetric. A convenient class of long-range
potentials are twice continuously differentiable functions with
$\,V^\ell(\bx)\to 0\,\text{ as }\,|\bx|\to\infty\,$ and
\begin{displaymath}
\mathcal{V}^\ell = \Bigl\{V^\ell \Bigm| |\nabla V^\ell(\bx)|\leq \const
(1+|\bx|)^{-(3/2)-\varepsilon},\; 
\end{displaymath}
\vspace*{-6ex}

\beql{10.1}
\qquad\qquad\quad\: |\Delta V^\ell(\bx)|\leq \const
(1+|\bx|)^{-2-\varepsilon},\;\varepsilon > 0 \Bigr\}.
\end{equation}
Local singularities may be treated with the short-range part.
If long-range potentials are present the ordinary scattering
operator (\ref{5.2}), (\ref{5.3}) no longer exits, the free
time evolution has to be replaced by a better approximation
generated by $\,H_0(t):=H_0 + V^\ell(t\,\bp/m)$. The factor
$\,e^{-it_{_-}H_0}\,$ in (\ref{5.2}) has to be replaced by
\beql{10.2}
U^D(t_{_-},0):=\exp \left\{-i\,t_{_-}\,H_0 -i\int_{t_{_-}}^0 dt
\;V^\ell(t\,\bp/m)\right\}
\end{equation}
and similarly $\,U^D(0,t_{_+})\,$ for the term with $\,t_{_+}\,$.
With this \textit{Dollard correction} the modified scattering
operator exists:
\beql{10.3}
S^D(t_{_+},t_{_-}) := 
[\,U^D(0,t_{_+})\,]^\ast \; e^{-i\, t_{_+} [H_0 + V^s + V^\ell]} \;
e^{i\, t_{_-} [H_0 + V^s + V^\ell]} \; U^D(t_{_-},0)\; ,
\end{equation}
\vspace*{-6ex}

\beql{10.4}
S^D := \spmlim 
\; S^D(t_{_+},t_{_-}), \quad\text{satisfies}\quad S^D \, \Phi^- =
\Phi^+.
\end{equation}
The splitting of the potential into its short- and long-range parts
is not unique. Different choices -- as long as (\ref{4.2}) and
(\ref{10.1}) are satisfied -- correspond to different labellings of
asymptotic configurations and are physically equivalent.

To define a modified scattering operator some knowledge of the
long-range behavior of the potential is needed. A typical physical
situation is that the total charge is known but the charge
distribution is to be determined by scattering experiments.

\begin{theorem}
{\bf (nonrelativistic, long-range, \cite{EnssWeder:Geometrical}) }\\
For a splitting of the potential as given above let the Hamiltonian be
\begin{displaymath}
H = \frac{1}{2m} \bp^2 + V^s(\bx) + V^\ell(\bx)
\end{displaymath}
and let $\,S^D\,$ be the corresponding Dollard modified scattering
operator then
\beql{10.5}
\begin{split}
v(\pb) \; &(\Phi'_\pb , \; i(S^D-\1)\, \Phi_\pb) - \int dr \; 
(\Phi'_\bn , \; [\, V^\ell(\bx + \om r) - V^\ell(\om r)\,]\; \Phi_\bn)\\[1ex]
& \pblim  \int dr \; (\Phi'_\bn , \; V^s(\bx + \om r)
 \, \Phi_\bn).
\end{split}
\end{equation}
\end{theorem}
Note that the integral $\,\int dr \;[\, V^\ell(\bx + \om r) -
V^\ell(\om r)\,]\,$ converges for every $\,\bx$.
Another version of the theorem yields at once the full potential
by using the high-energy limit of $\,[p_1,\;S^D]\,$ for \textit{any}
modified scattering operator.

If one has a magnetic field described by a vector potential $\,\A\,$
in a nonrelativistic Schr\"odinger operator then the leading term in
the interaction is $\,\A\!\cdot\!\bp/m \approx \A\!\cdot\!\bv(\pb)$.
Inserting this into (\ref{8.8}) suggests that one obtains a
nontrivial limit of the scattering operator although the kinematics
is nonrelativistic. The other potentials show up as a correction to
the leading behavior of lower order.
\begin{theorem}
{\bf (nonrelativistic with magnetic fields and potentials of short
and long range, \cite{Nicoleau}, \cite{Arians1},
\cite{Arians3})}\\ Let the vector potential $\,\A(\bx)\,$ be a
short-range multiplication operator and
\beql{10.6}
\begin{aligned}
H &= \frac{1}{2m}\: [\,\bp - \A(\bx)\,]^2 + V^s(\bx) + V^\ell(\bx) \\[0.9ex]
  &= \frac{1}{2m} \bp^2 - \A\cdot \frac{\bp}{m} + 
\frac{1}{2m}(i\, \Div \A + |\A|^2) + V^s + V^\ell .
\end{aligned}
\end{equation}
The scattering operator itself has the limit
\beql{10.7}
\slim_{|\pb| \to \infty} e^{-i\,\pb\bx} S  e^{i\,\pb\bx} = 
\exp \left\{ i\int dr\; \om\cdot\A(\bx + \om r) \right\}.
\end{equation}
which yields a unique magnetic field $\,F(\bx)=d\A(\bx)\,$ for continuous 
$\,\A(\bx)$, and -- after fixing a gauge -- the 
vector potential $\,\A(\bx)$. In a second step one obtains as a
lower order correction to the leading term
\beql{10.8}
\begin{split}
v(\pb)\! & \left( \Phi'_\pb , \; i\!\left[ S \: e^{- i\int dr\; 
\om\cdot\A(\bx + \om r) } - \1 \right] \Phi_\pb \right) - 
\mathrm{known \; terms}(\A, V^\ell,\, \Phi'_\bn ,\, \Phi_\bn)\\[1ex]
& \pblim  \int dr \; 
(\Phi'_\bn , \;V^s(\bx + \om r) \; \Phi_\bn).\\
\end{split}
\end{equation}
\end{theorem}
Clearly, only the magnetic field $\,d\A(\bx)\;(= \text{curl }A(\bx)\,$
in$\,\nu=3$ dimensions) is a physical quantity. However, fixing a gauge and
choosing a vector potential is needed to write down the
Schr\"odinger operator (\ref{10.6}) (choice of a representation).
For homogeneous magnetic fields see \cite{Arians2}, \cite{Arians3}.

Let us now turn to nonrelativistic {\bfseries multiparticle} scattering 
systems with short- and long-range potentials. As the interaction becomes
weak for high speeds the incoming and outgoing scattering channels are
the same in the high-energy limit. Excitations and rearrangements
are of lower order. 

\begin{theorem}
{\bf (nonrelativistic multiparticle systems)}\\
In two-cluster scattering the limit of the channel scattering operator 
for high relative speed of the clusters yields the effective intercluster 
potential between the two bounded subsystems, 
{\em \cite{EnssWeder:TwoCluster}}.

In the totally free channel select a pair of particles with high relative 
speed and all other particles far away (or with even higher speeds). Then 
the pair potential of this pair is obtained from the limit. Multiparticle 
potentials can be obtained iteratively after the pair potentials are known, 
{\em \cite{EnssWeder:Uniqueness}, \cite{EnssWeder:Geometrical}}.
\end{theorem}

The method has been extended and applied to other problems as well, see
e.g.\
Stark effect Hamiltonians \cite{Weder:Electric},
time-dependent $N$-body Schr\"odinger operators \cite{Weder:TimeDep},
time-dependent Dirac operators \cite{Ito:TimeDep}, 
and
nonlinear Schr\"odinger equations \cite{Weder:NLS}.

% section 11 11 11 

\section{Reconstruction of the Potential} \label{SecReconstruction}

The condition $\,\nu \geq 2\,$ (\textit{multi}dimensional inverse problem)
enters here to obtain from the above limits reconstruction formulas and
uniqueness.
For bounded continuous (or more general) functions $\,V^s\,$ the expression
\beql{11.1}
W^s(\bx, \om) := \int dr \; V^s(\bx + \om r)
\end{equation}
is the X-ray transform of $\,V^s$. In $\,\nu = 2\,$ dimensions lines
and hyperplanes are the same. Therefore, (\ref{11.1}) is the Radon
transform as well. The latter is known to be uniquely invertible because the 
assumption (\ref{4.2}) implies $\,V^s \in L^2(\R^2)$, see e.g.\
Theorem 2.17 in Chapter I of \cite{Helga}.
The inverse Radon transform yields the unique potential.
In higher dimensions one fixes e.g.\ $\,x_3,\, \dots,\, x_\nu\,$ and
reconstructs the ``slices'' subsequently.  In particular,
it is sufficient to vary $\,\om\,$ in a two dimensional plane. For unbounded
or discontinuous potentials the expectation value between states from
a dense set of nice vectors (like those which satisfy (\ref{3.7})\,)
effectively smoothes the potential. This is enough to reconstruct the
potential as a multiplication operator, see e.g.\ \cite{EnssWeder:Geometrical}.

The reconstruction of the magnetic field is similar but more complicated, 
see \cite{Arians1}, \cite{Arians3}, or \cite{Jung} for details.

\end{document}